\documentclass[fleqn,usenatbib]{mnras}

\usepackage{times}

\usepackage[T1]{fontenc}
\usepackage{ae,aecompl}
\usepackage[usenames,dvipsnames]{xcolor}
\usepackage[normalem]{ulem}

\usepackage{graphicx, amssymb, aas_macros} 
\newcommand{\mvir}{M_{\rm{vir}}}

\newcommand{\rvir}{R_{\rm{vir}}}

\newcommand{\mstar}{M_{\star}}
\newcommand{\msun}{M_{\odot}}

\newcommand{\mpc}{{\rm Mpc}}

\newcommand{\lcdm}{$\Lambda$CDM}

\newcommand{\hst}{\textit{HST}}
\newcommand{\jwst}{\textit{JWST}}

\def\nar{{New~Astron.~Rev.}}  

\newcommand{\protoLG}{proto-Local-Group}
\newcommand{\scriptV}{\mathcal{V}}

\title[The Local Group: The Ultimate Deep Field] {The Local Group: The Ultimate Deep Field}
\author[Boylan-Kolchin et al.]  {Michael
  Boylan-Kolchin$^1$\thanks{$\!$mbk@astro.as.utexas.edu},
  Daniel R. Weisz$^{2}$, 
  James S. Bullock$^3$ 
\newauthor 
  and Michael C. Cooper$^3$\\
\noindent $\!\!^1$Department of Astronomy, The University of Texas at Austin,
2515 Speedway, Stop C1400, Austin, TX 78712-1205, USA\\
\noindent $\!\!^2$Department of Astronomy, University of California Berkeley, Berkeley, CA 94720, USA\\
\noindent $\!\!^3$Department of Physics and Astronomy, University of California
at Irvine, Irvine, CA 92697, USA}
\date{\today}
\pubyear{2016}
\begin{document}
\label{firstpage}
\pagerange{\pageref{firstpage}--\pageref{lastpage}}
\maketitle

\begin{abstract}
  \textit{Near-field cosmology} -- using detailed observations of the Local
  Group and its environs to study wide-ranging questions in galaxy formation and
  dark matter physics -- has become a mature and rich field over the past
  decade.  There are lingering concerns, however, that the relatively small size
  of the present-day Local Group ($\sim 2\,\mpc$ diameter) imposes
  insurmountable sample-variance uncertainties, limiting its broader utility. We
  consider the region spanned by the Local Group's progenitors at earlier times
  and show that it reaches $3' \approx 7$ co-moving Mpc in linear size (a volume
  of $\approx 350\,\mpc^3$) at $z=7$. This size at early cosmic epochs is large
  enough to be representative in terms of the matter density and counts of dark
  matter halos with $\mvir(z=7) \la 2\times 10^{9}\,\msun$. The Local Group's
  stellar fossil record traces the cosmic evolution of galaxies with
  $10^{3} \la \mstar(z=0) / \msun \la 10^{9}$ (reaching $M_{1500} > -9$ at
  $z\sim7$) over a region that is comparable to or larger than the
  \textit{Hubble} Ultra-Deep Field (HUDF) for the entire history of the
  Universe. It is highly complementary to the HUDF, as it probes \textit{much}
  fainter galaxies but does not contain the intrinsically rarer, brighter
  sources that are detectable in the HUDF. Archaeological studies in the Local
  Group also provide the ability to trace the evolution of individual galaxies
  across time as opposed to evaluating statistical connections between
  temporally distinct populations.  In the \jwst\ era, resolved stellar
  populations will probe regions larger than the HUDF and any deep \jwst\
  fields, further enhancing the value of near-field cosmology.
\end{abstract}


\section{Introduction} 
The standard introduction of a paper on near-field cosmology extols the virtues
of the Local Group as a cosmic Rosetta Stone that provides archaeological clues
left behind by untold generations of stars, clues that may unlock unsolved
mysteries in galaxy formation. A frequent concern is that resolved-star studies
are inherently limited to a region that is relatively small and likely biased,
however, setting fundamental limits on the broader applicability of results
based on near-field studies.  In this \textit{Letter}, we show that the volume
spanned by the high-redshift progenitors of the Local Group was large enough to
have been typical in many respects. The archaeological record imprinted on Local
Group galaxies is therefore likely to provide an unbiased view of faint galaxy
populations at early times, making near-field observations a powerful complement
to direct deep-field studies.

\section{Methods}
\label{section:methods}
At present, the Local Group consists of two dark matter halos, each with virial
mass of $\sim 10^{12}\,\msun$ (e.g., \citealt{klypin2002}), approaching each
other for presumably the first time \citep{kahn1959}. The co-moving Lagrangian
volume of the matter contained within the Local Group -- i.e., the co-moving
volume spanned at earlier epochs by the particles in the current-day Local Group
-- was therefore larger in the past; the same is true for all dark matter
halos. Thus, while the Local Group is a very specific (and non-typical) volume
of the Universe today, its properties at earlier times should be closer to an
average portion of the Universe. Precisely how much closer is the topic of this
paper.

The spherical collapse model \citep{gunn1972} provides a starting point for
understanding the evolution of the Local Group. In this model, the matter within
a virialized dark matter halo of radius $\rvir$ (defined by an average enclosed
density of $\Delta_{\rm vir}$ times the critical density of the Universe) came
from a spherical region with a co-moving Lagrangian radius $r_{l}$ equal to
$(\Delta_{\rm vir}/\Omega_{\rm m})^{1/3}\,\rvir$.  For flat cosmologies with
$\Omega_{\rm m} \sim 0.3$, this gives $r_l \approx 7 \, \rvir$, indicating that
the Local Group must have been significantly larger in co-moving size in the
early Universe. Cosmological zoom-in simulations, by design, follow the
evolution of Lagrangian regions surrounding specific halos from linear
fluctuations to the highly non-linear regime \citep{katz1993, onorbe2014}. The
ELVIS suite of $N$-body zoom-in simulations \citep{garrison-kimmel2014} provides
12 Local Group analogs simulated from $z=125$ to $z=0$, each of which is
uncontaminated by lower resolution particles over a spherical region having a
radius of at least $1.2\,\mpc$ centered on the $z=0$ barycenter of the Local
Group. In what follows, we use this suite to study the co-moving volume probed
by the Local Group at higher redshifts.

In our analysis, we first eliminate three ELVIS pairs that contain a third
large, nearby halo, as these would bias any results. For the remaining 9 pairs,
we identify all subhalos within $1.2\,\mpc$ of the Local Group's $z=0$
barycenter and track \textit{all} of their progenitors back through time. There
are occasionally individual subhalos that come from regions that are distant
from the vast majority of the matter that forms the Local Group; such subhalos
can artificially increase the inferred volume of the Local Group at earlier
times. To eliminate these objects, we run a friends-of-friends \citep{davis1985}
group finder with a large linking length of 400 kpc and retain only the main
grouping. In practice, this removes $\ll 1\%$ of subhalos at $z=0$. We then
identify the positions spanned by the progenitors of the remaining Local Group
subhalos above the ELVIS completeness limit of
$M_{\rm peak}=6 \times 10^7\,\msun$ at each earlier snapshot; this constitutes
the ``proto-Local-Group'' at each epoch. It is important to note that the number
of galaxies or halos in the \protoLG\ is much larger than the number in the
Local Group at $z=0$ owing to mergers and disruption over time.

If we are only interested in understanding the Local Group itself, this would be
sufficient. To place the Local Group in context at higher redshifts, however, we
must understand the full environment that the \protoLG\ occupies. We therefore
compute, at each snapshot, the minimum cuboid volume defined by the \protoLG\ --
i.e., the rectangular cuboid defined by the minimum and maximum co-moving
coordinate locations of all \protoLG\ progenitors at that time,
$\scriptV_{\rm RC}$ -- and identify all additional halos in this region (i.e.,
halos that appear to be part of the \protoLG\ but that do not end up in the
Local Group at $z=0$). The inclusion of these objects roughly doubles the counts
at $z \sim 7$ within $\scriptV_{\rm RC}$. This doubling is consistent with the
extra volume contained in the cuboid $\scriptV_{\rm RC}$ region circumscribing a
sphere, although the \protoLG\ progenitors are not confined to a spherical
region at $z \sim 7$. Proto-Local-Group halos dominate the central portion of
$\scriptV_{\rm RC}$ and the additional halos populate the outskirts of the
volume.

We define the linear size of the \protoLG\, $l_{\rm LG}(z)$ as the geometric
mean of the three axes defining $\scriptV_{\rm RC}(z)$; in other words,
$l_{\rm LG}(z)=\scriptV_{\rm RC}(z)^{1/3}$. At $z=0$, the Local Group volume is
defined by a sphere of radius $1.2\,\mpc$, so
$l_{\rm LG}(z=0) \approx 2.4\,\mpc$ (the actual number depends on the
distribution of halos at $z=0$ but can never exceed $2.4\,\mpc$). At higher
redshifts, $\scriptV_{\rm RC}$ can, in principle, become highly elongated in one
or two dimensions. In practice, however, we find that this is not the case: at
$z=7$, the median minor-to-major axis ratio is 0.76 and the median
intermediate-to-major axis ratio is 0.81, and in only one case is the minor axis
smaller than half of the major axis size. The typical $\scriptV_{\rm RC}(z=7)$
is moderately prolate: 6 of 9 simulated \protoLG s have a triaxiality parameter
$T$ (see \citealt{franx1991}) larger than 0.5.

\section{The Local Group Through Time}
\label{sec:LGthruT}
\begin{figure}
\centering
\includegraphics[width=0.45\textwidth]{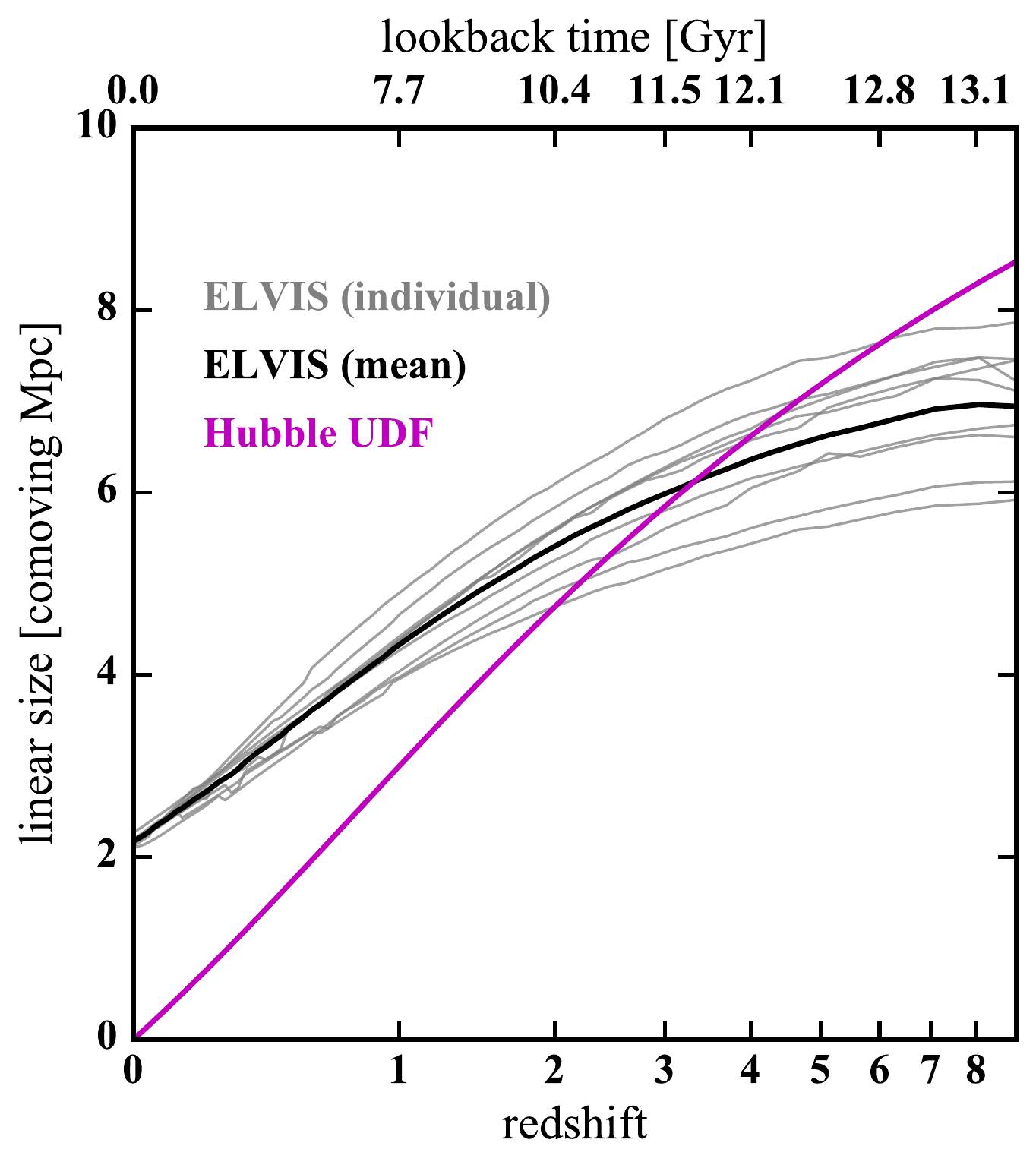}
\caption{The co-moving linear extent of the \protoLG\ (black, gray curves) and HUDF
  (magenta curve) as a function of redshift. For $z\la3$, the \protoLG\ covers an
  area on the sky that is larger than the HUDF. 
  At earlier 
  times, the HUDF is marginally larger. The typical \protoLG\ reaches a co-moving
  size of $7\,\mpc$ at $z\sim 7$, meaning it probes an effective volume of $\sim
  \! 350$ co-moving $\mpc^3$ in the reionization era.
\label{fig:lg_udf}
}
\end{figure}

Figure~\ref{fig:lg_udf} shows the co-moving linear size, $l_{\rm LG}(z)$, of the
\protoLG\ going back in time to $z=9$. Thin gray lines show the size of
individual Local Group pairs from the ELVIS simulation suite, while the thick black line
shows the median value across the ELVIS pairs at each redshift. The linear size
of the \protoLG\ increases with increasing redshift, reaching $\approx 7\,\mpc$
(co-moving) at $z\sim 7$. Going back in time, therefore, the Local Group probes
a significantly larger (co-moving) volume than it does today. To give context to
the Local Group's size at earlier epochs, Fig.~\ref{fig:lg_udf} also shows the
co-moving linear size of the HUDF (\citealt{beckwith2006}, assuming an angular
size of $3.1' \times 3.1'$) as a function of redshift (magenta
curve). \textit{At all epochs later than $z\approx 3$ (the last 85\% of cosmic
  time), the proto-Local-Group covers a larger area on the sky than the HUDF.}

\begin{figure}
\centering
\includegraphics[width=0.45\textwidth]{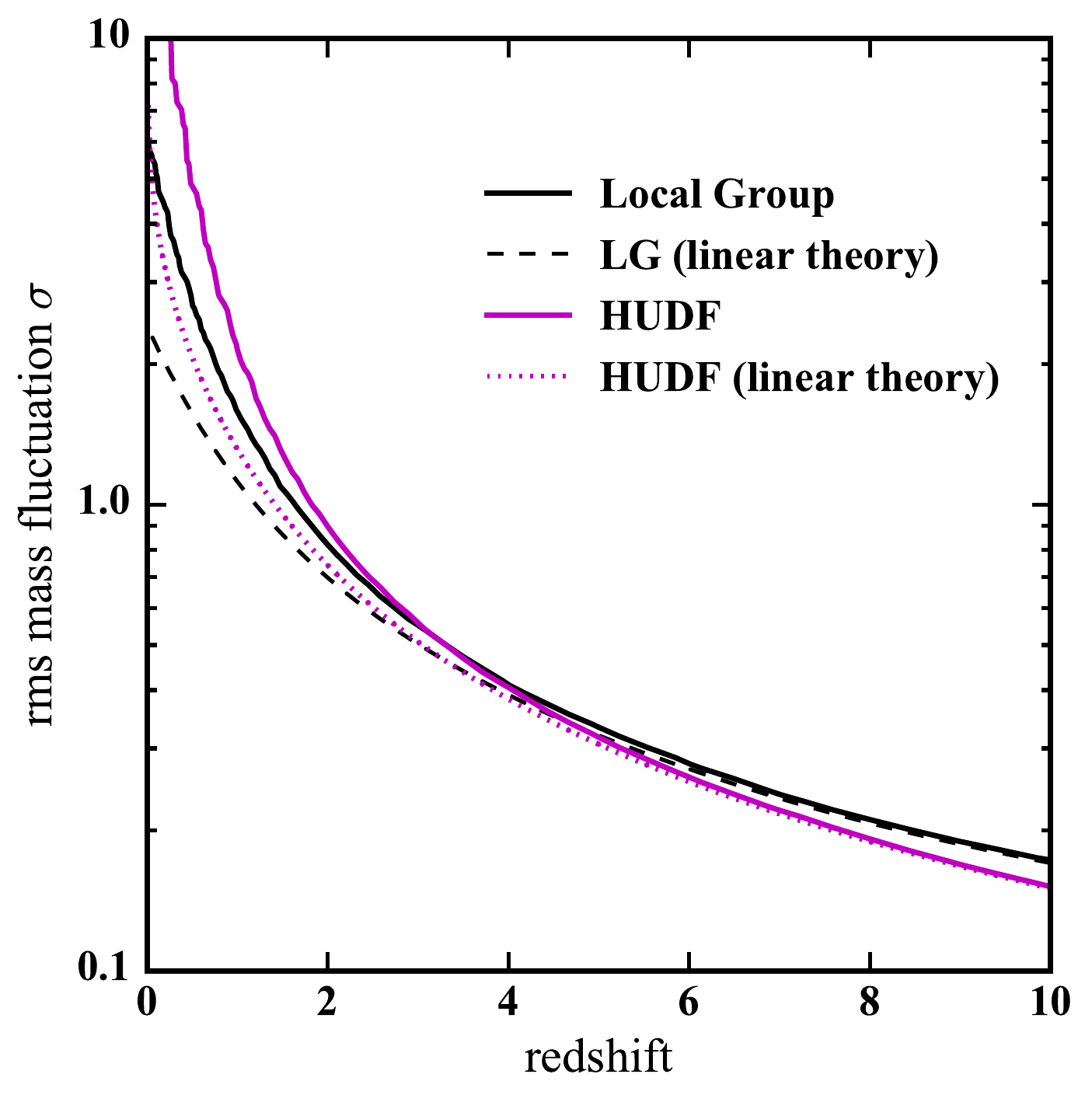}
\caption{The rms mass fluctuation, $\sigma$, computed over volumes equal to that
  covered by the \protoLG\ (black) and HUDF (magenta) as a function of
  redshift. Also shown are linear theory values for $\sigma$ over those same
  volumes. At early times, both regions have $\sigma \sim 0.2$ and the redshift
  evolution is well-described by linear theory. The regions beome non-linear
  ($\sigma \ga 1$) at $z \sim 2$, by which point linear theory underestimates the
  true rms amplitude of fluctuations.
\label{fig:lg_rmsfluc}
}
\end{figure}

It is important to understand how representative such portions of the Universe
are at each cosmological epoch. One way to do this is to compute the rms
amplitude of density fluctuations $\sigma$ in regions having volumes equal to
$\scriptV_{\rm RC}(z)$. In classical Press-Schechter \citeyearpar{press1974}
theory and its extensions, the typical scale $M^*$ that is collapsing at a given
epoch has $\sigma_{\rm lin}(M^*, z)=\delta_{\rm c}\approx 1.686$ (where
subscript ``lin'' indicates that the relevant rms amplitude comes from linear
theory, extrapolated to the redshift in question). Roughly speaking, scales with
$\sigma(M, z) \gg 1$ have collapsed while those with $\sigma(M, z) \ll 1$ are
firmly in the linear regime.

In reality, linear theory underestimates $\sigma(M,z)$.  Cosmological
simulations account for effects of non-linear growth; accordingly, we use the
{\it Illustris} suite \citep{vogelsberger2014, vogelsberger2014a, nelson2015} to
compute $\sigma(M,z)$. At each snapshot, we evaluate the density field for
Illustris-Dark-1 on a $256^3$ grid, compute the overdensity
$\delta_i=\rho_i/\langle \rho \rangle-1$ for each cell $i$, then calculate
$\delta(M,z)$ by smoothing the gridded overdensity field with a real-space
top-hat filter having a volume equal to $\scriptV_{\rm RC}(z)$ (so the mass
contained in the volume is $M_{\rm LG}=\rho_{\rm m}(z)\,\scriptV_{\rm RC}(z)$).
The rms amplitude of fluctuations is equivalently characterized by
$\sigma(M_{\rm LG},z)$ or $\sigma(l_{\rm LG},z)$

The resulting values for $\sigma(l_{\rm LG},z)$ are plotted as a solid black
curve in Figure~\ref{fig:lg_rmsfluc}; the linear theory value of
$\sigma(l_{\rm LG},z)$ is shown as a dashed black curve. We also compute the
same quantities for the HUDF, $\sigma(l_{\rm HDF},z)$, and plot these with
magenta curves. Both the Local Group and the HUDF have $\sigma < 1$ for
$z \ga 2$. At high redshift, both probe volumes that are well-described by
linear theory. In particular, the volumes probed by the \protoLG\ and by cubic
slices of the HUDF ($\Delta\,z \approx 0.02$) in the reionization era
($6 \la z \la 10$) have $\sigma(M_{\rm LG}) \la 0.25$. For a broad discussion of
variance in deep-field galaxy counts, see \citet{robertson2010a}.

\begin{figure}
\centering
\includegraphics[width=0.495\textwidth]{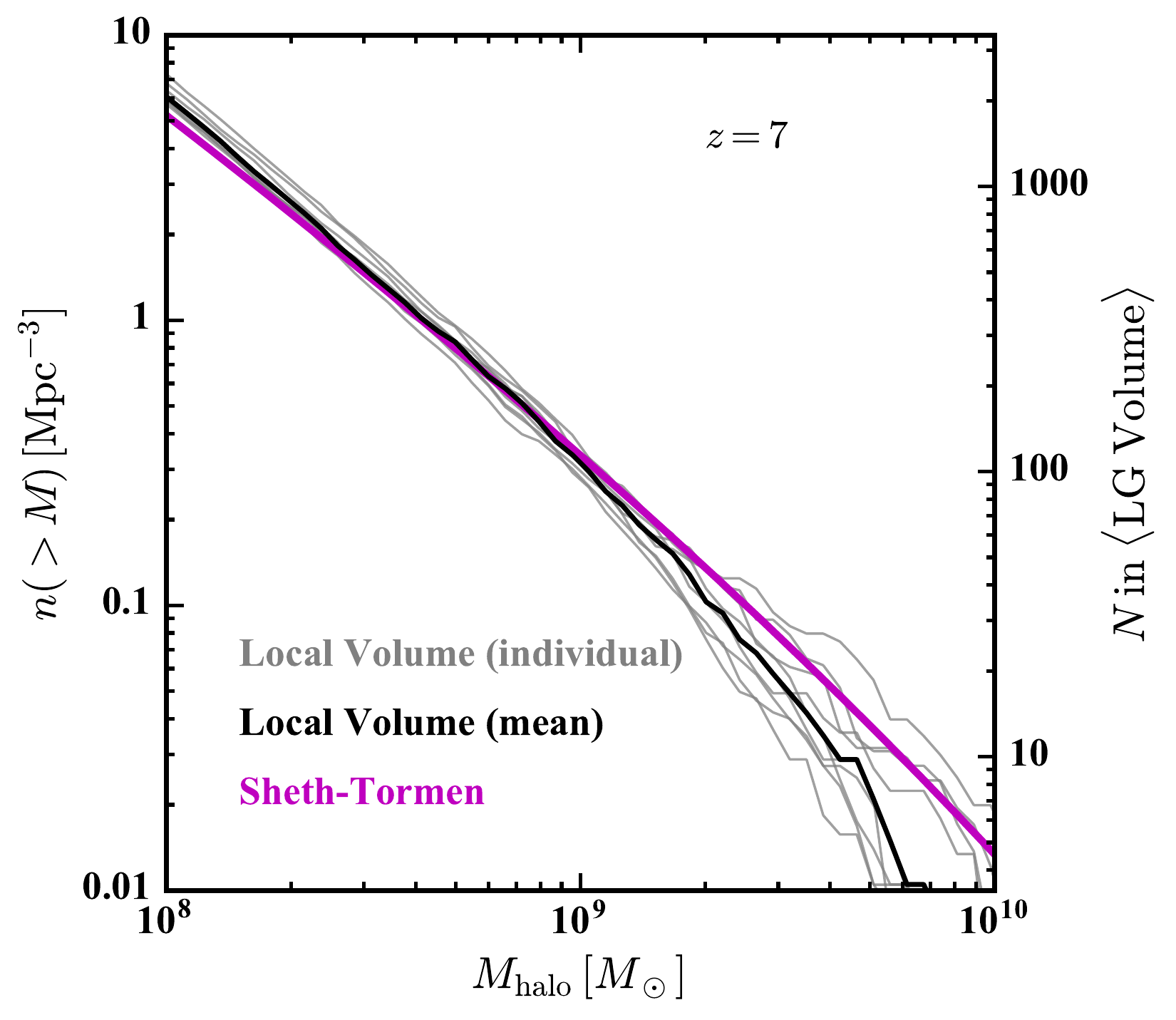}
\caption{Cumulative dark matter halo mass functions for the volume spanned by
  the \protoLG\ at $z=7$ (gray curves: individual ELVIS simulations; black curve:
  ensemble average). The right y-axis gives the cumulative counts for
  the average \protoLG\ region in ELVIS. The magenta curve shows the global
  Sheth-Tormen mass function. At $z=7$, the normalization of the ELVIS \protoLG\
  mass function agrees very well with the Sheth-Tormen mass function, indicating
  the \protoLG\ covers a representative portion of the Universe.
\label{fig:lg_massfn_sheth-tormen}
}
\end{figure}

The results above have established that the \protoLG\ was substantially larger
at earlier epochs, large enough to cover a volume that becomes non-linear only
after $z \sim 2$. A further, and more stringent, test is to compare the mass
function in the region defined by the \protoLG\ to the cosmological mass
function at earlier times. In Figure~\ref{fig:lg_massfn_sheth-tormen}, we plot
the cumulative co-moving number density $n(>M)$ of halos within
$\scriptV_{\rm RC}$ for each ELVIS pair (thin gray lines), as well as the median
across the simulation suite (thick black line), at $z=7$. The cosmological
expectation, as encapsulated by the Sheth-Tormen \citeyearpar{sheth2001} mass
function, is plotted as a magenta line. The mass function in
$\scriptV_{\rm RC}(z=7)$ matches the cosmological mass function for
$M_{\rm vir} \la 2\times 10^{9}\,\msun$, with counts in the \protoLG\ region
falling below Sheth-Tormen at higher masses (smaller number densities) owing to
the size of this region. \textit{The volume covered by the \protoLG\ at $z=7$ is
  therefore a cosmologically representative region for the mass function of
  halos with $\mvir(z=7) \la 2 \times 10^{9}\,\msun$}, a remarkable result.

As indicated by the scale on the right side of the figure, there should be
$\sim 2000$ halos with $\mvir > 10^8\,\msun$ (this approximately corresponds to
halos above the atomic cooling threshold) and $\sim 100$ halos with
$\mvir > 10^9\,\msun$ in the $z=7$ \protoLG\ region. It is this large number of
low-mass systems, coupled with the small value of $\sigma(l_{\rm LG}, z=7)$,
that makes mass functions in the \protoLG\ cosmologically representative even in
the $350\,\mpc^3$ co-moving volume at early times (note the small variance in
normalizations of the mass functions in Fig.~\ref{fig:lg_massfn_sheth-tormen}).

\begin{figure*}
\centering
\includegraphics[width=0.95\textwidth]{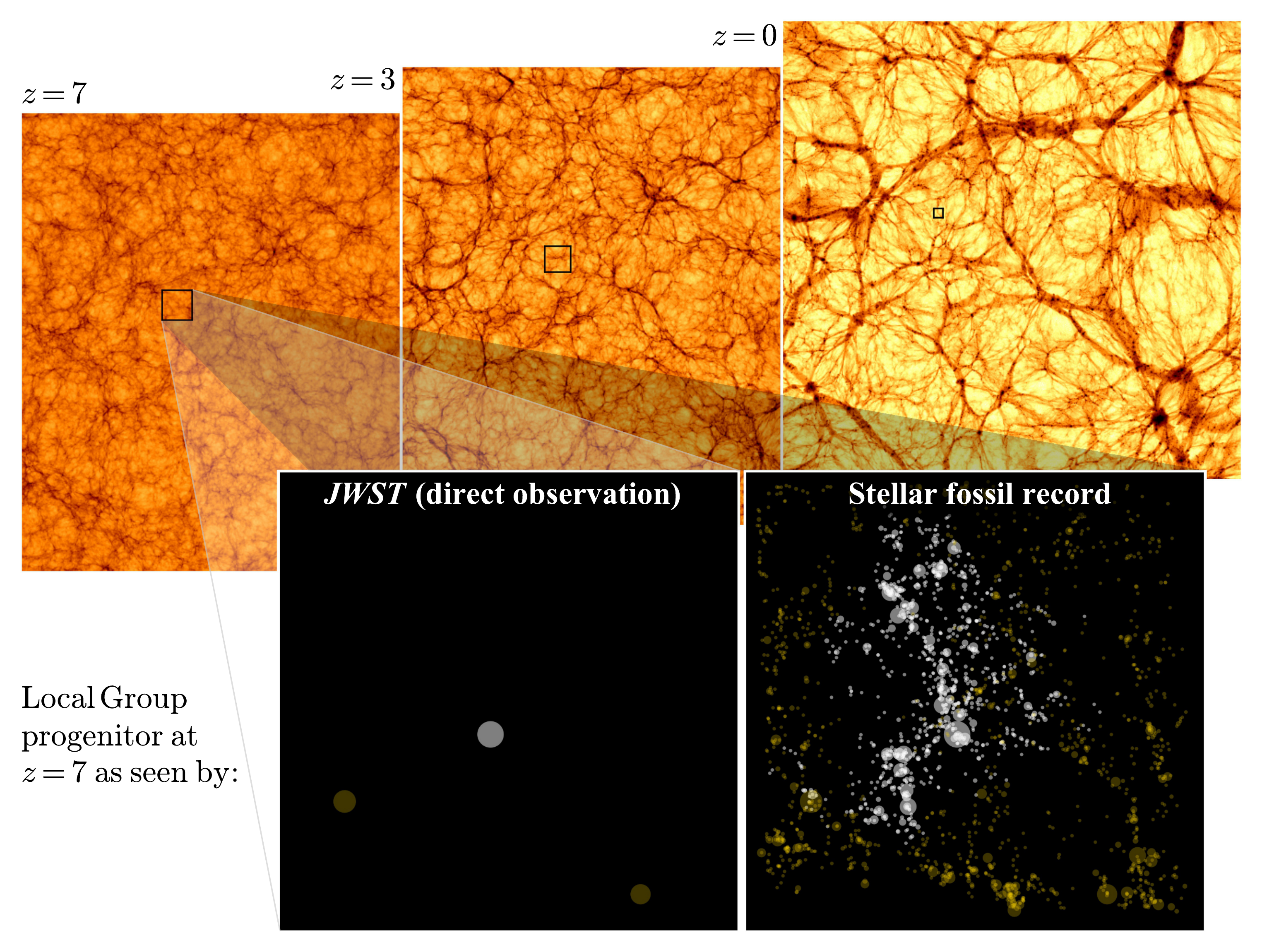}
\caption{Thin density slices of the {\it Illustris} simulation at
  $z=0,\,3,\,{\rm and}\;7$ (upper panels; $106.5 \times 106.5 \times 1.065$
  co-moving Mpc), along with the co-moving size of the average \protoLG\ at each
  redshift (top panels). The bottom panels show the galaxies in such a region
  (which is comparable in linear extent to the HUDF) that are accessible at
  $z=7$ through direct observation with \jwst\ in a narrow redshift slice of
  width $\Delta z \approx 0.02$ (corresponding to the $\sim 7$ Mpc co-moving
  size of the \protoLG\ at $z = 7$) and through the stellar fossil record in the
  Local Group. White symbols indicate objects that end up in the Local Group at
  $z=0$, while gold symbols represent objects in the \protoLG\ volume at $z=7$
  but not within $1.2\,\mpc$ of the Local Group barycenter at $z=0$.
\label{fig:composite_img}
}
\end{figure*}

\begin{figure}
\centering
\includegraphics[width=0.47\textwidth]{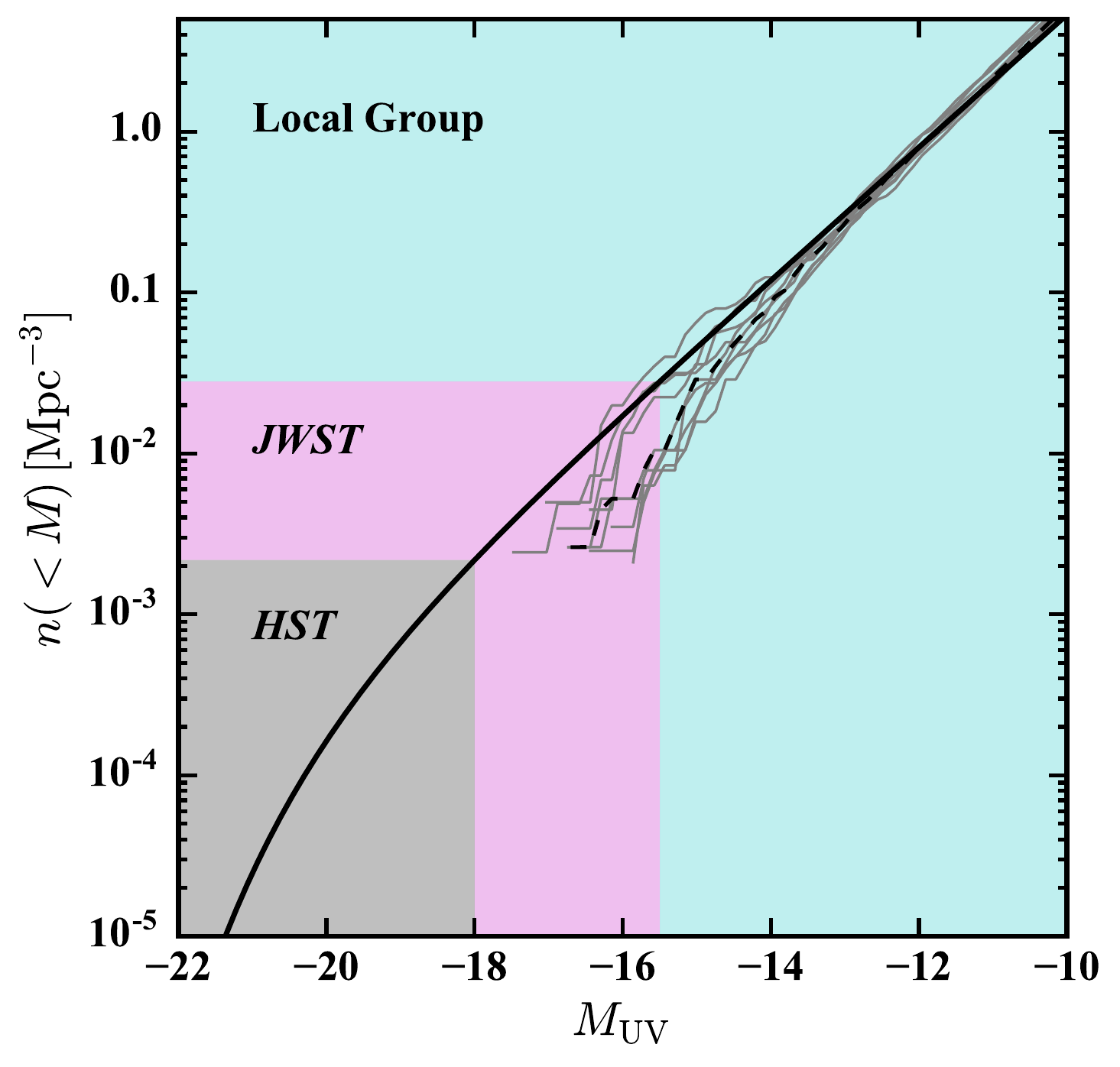}
\caption{ \label{fig:uvlf} Complementarity of direct observation and stellar
  fossil record studies of galaxies at high-$z$. \textit{Solid line:} cumulative
  luminosity function at $z \sim 7$ from deep-field observations using \hst\
  (from \citealt{finkelstein2015}). The best-fitting cumulative luminosity
  function is extrapolated beyond the HUDF magnitude limit of $M_{\rm UV}=-18$. 
  \textit{Solid gray curves} show the cumulative luminosity functions
  found in each ELVIS pair based on abundance matching. The agreement in
  normalization between the luminosity function globally and in the \protoLG\
  region is a direct result of the unbiased nature of counts in the \protoLG\
  volume at $z=7$ (Fig.~\ref{fig:lg_massfn_sheth-tormen}).  The approximate
  blank field limiting magnitude for \jwst\ is also shown, along with the
  corresponding number density. The Local Group probes galaxies that, for the
  most part, will remain undetected through direct observations in the high-$z$
  Universe because of their intrinsic faintness.}
\end{figure}
 By $z=4$, the \protoLG\ has become somewhat more dense, and the equivalent plot
of mass functions shows that counts in the \protoLG\ volume exceed Sheth-Tormen
expectations by $\approx 50\%$ at this epoch. However, the shape of the mass
functions still matches that of Sheth-Tormen for cumulative number densities
larger than $10^{-2}\,\mpc^{-3}$ (equivalently, $\mvir < 5 \times
10^{10}\,\msun$).

\section{Discussion}
\label{sec:disc}
The previous sections have explored a simple yet important question: given
regions that end up as Local Group analogs at $z=0$, how representative is the
volume that their progenitors spanned at earlier times? The rough estimate given
at the start of Sec.~\ref{section:methods} indicates that a collapsed dark
matter halo covered a comoving region that was $\ga 7$ times larger in the early
Universe than at $z=0$, and the results of Sec.~\ref{sec:LGthruT} show that the
Local Group (which is not a virialized region today) was approximately 3.5 times
larger than its present-day size at early times. The \protoLG\ is large enough
be representative at high redshifts both for the matter density it contains and
for counts of dark matter halos with $\mvir \la 2\times 10^9\,\msun$.

These results are only true when starting with Local-Group-like regions at $z=0$
and considering their properties at earlier times; \protoLG-sized regions
selected at $z\sim7$ are large enough to be representative at that time but will
not generally evolve to be Local Groups at $z=0$. The direct, one-to-one
connection between galaxies in the $z=0$ Local Group and their ancestors in the
\protoLG\ is complicated by any mergers and disruption of galaxies in the
intervening time; this also means that surviving galaxies at $z=0$ represent a
lower limit on the number of similar galaxies in the \protoLG. While we plan to
address this point in future work, we note that the predicted merger histories
of low-mass halos in \lcdm\ are such that most present-day dwarfs have not had
significant mergers (in terms of stellar mass growth) since the reionization era
\citep{deason2014a, boylan-kolchin2015}; Local Group dwarfs are therefore
expected to provide a direct window to low-mass systems in the high-redshift
Universe.

The representative nature of the \protoLG's volume, coupled with the similarity
between the (1D) size of the \protoLG\ and the HUDF, suggest an interpretation
of the Local Group at eariler epochs: \textit{observations of the Local Group
  can be thought of as providing a (very) narrow slice in time of the HUDF.}
Given our ability to measure resolved star formation histories of Local Group
galaxies, we can look at the Local Group at a variety of ``snapshots'' in
time. This is the same as looking at a series of thin transverse
slices through the HUDF. A complete census of galaxies within 1--2 Mpc of the
Local Group with depth to reach the oldest main sequence turn-off would
therefore allow a \textit{continuous} look at galaxy formation and evolution --
i.e., tracking individual galaxies across cosmic time -- in a size equivalent to
the HUDF to depths of $m \sim 38$ ($M_{\rm UV} \approx -9$) at $z=7$.

We emphasize that resolved-star studies of the Local Group provide an almost
perfectly complementary view of galaxy formation to deep blank-field (and
lensing) observations of the high-redshift Universe with \textit{Hubble}. At
$z\sim 7$, the faintest galaxies in blank-field \textit{HST} observations are
likely to be more massive than the progenitor of the Milky Way at that time
\citep{boylan-kolchin2014}. Archaeological studies of Local Group galaxies
extend the range of galaxies to at least 8 magnitudes fainter \citep{weisz2014c,
  boylan-kolchin2015, graus2016}. \textit{HST} is therefore capable of probing
galaxy formation over six decades in mass ($10^{3} < \mstar/\msun < 10^{9}$) and
12--13 Gyr in time over an area comparable to the HUDF via the stellar fossil
record.

The power of near-field studies, and their complementarity to direct high-$z$
observations, is emphasized in Figures~\ref{fig:composite_img} and
\ref{fig:uvlf}. For both figures, we assign UV luminosities at $z=7$ to ELVIS
dark matter halos via abundance matching based on the global UV luminosity
function from \citet{finkelstein2015} and the Sheth-Tormen mass
function. Figure~\ref{fig:composite_img} shows slices through the density
distribution of the {\it Illustris} simulation at $z=0,\,3,\, {\rm and}\; 7$,
along with boxes indicating the approximate size of the \protoLG\ at each of
those redshifts. The insets at $z=7$ show galaxies that can be observed either
directly in the \protoLG\ with the \textit{James Webb Space Telescope} (\jwst;
left) or through archaeological studies in the Local Group with \hst\
(right). \jwst\ deep fields will have many more galaxies at a range of
redshifts, while the stellar fossil record in the Local Group probes a huge
number of galaxies that will be unobservably faint at cosmic dawn.

Fig.~\ref{fig:uvlf} shows the cumulative galaxy luminosity function from
\citet{finkelstein2015} at $z=7$ (thick black curve), extrapolated from the
\hst\ observational limit of $M_{\rm UV}\sim -18$ to 
$M_{\rm UV}=-10$. Each gray line shows the luminosity function in the \protoLG\
volume from an individual ELVIS simulation, while the dashed black line shows
the mean among all ELVIS pairs. Importantly, the agreement in normalization
between the global luminosity function and the \protoLG\ regions from ELVIS was
not pre-determined but rather is a result of the agreement between the global
halo mass function and the halo mass function within the \protoLG\ at $z=7$. The
approximate reach of \textit{HST} and \jwst\ is shown in the figure, while the
\protoLG\ covers the entire region of the figure in which the halo mass
functions are non-zero. 

The high angular resolution of \jwst\ and its exquisite sensitivity to (red)
ancient main sequence turn-off stars make it ideally suited to extend the same
census to 3--5 Mpc, meaning it is capable of surveying a region within the Local
Volume that is markedly larger than the HUDF (and any \jwst\ deep field) at all
redshifts. The complementarity with direct observations of galaxies at early
cosmic epochs will also be enhanced, as \jwst\ will likely detect galaxies as
faint as $M_{\rm UV}\sim -15.5$ in blank fields at $z\sim7$
\citep{windhorst2006}, comparable to progenitors of the Large Magellanic Cloud
(\citealt{boylan-kolchin2015}; see \citealt{patej2015} for related
calculations). The \textit{Wide-Field Infrared Survey Telescope}
(\textit{WFIRST}) has potential to be similarly transformative for near-field
cosmology. The addition of an optical filter (e.g., $R$-band) would improve its
angular resolution (and temperature sensitivity), allowing it to reach the
ancient main sequence turn-offs of galaxies out to $\sim$5 Mpc. This
modification, combined with \textit{WFIRST}'s extremely wide field of view,
would capture full star formation histories over the entire spatial extent of
virtually any nearby galaxy in a single pointing, revolutionizing how we study
and understand the evolution of low-mass galaxies.

\section*{Acknowledgments} 
We thank Mark Vogelsberger and Lars Hernquist for assistance with accessing and
analyzing {\it Illustris} data via the Odyssey cluster, which is supported by
the FAS Division of Science Research Computing Group at Harvard University. MBK
acknowledges support from NSF grant AST-1517226 and from NASA grants
HST-AR-12836 and HST-AR-13888 awarded by STScI. JSB was supported by
NSF-AST-1518291, HST-AR-14282, and HST-AR-13888. Much of the analysis in this
paper relied on the python packages {\tt NumPy} \citep{numpy}, {\tt SciPy}
\citep{scipy}, {\tt Matplotlib} \citep{matplotlib}, and {\tt iPython}
\citep{ipython}, as well as {\tt pyfof}
(\href{https://pypi.python.org/pypi/pyfof/}
{https://pypi.python.org/pypi/pyfof/}); we are very grateful to the developers
of these tools. This research has made extensive use of NASA's Astrophysics Data
System and the arXiv eprint service at arxiv.org.

\bibliography{draft.bbl}
\label{lastpage}
\end{document}